\documentclass[lettersize,journal]{IEEEtran}
\usepackage{amsmath}
\usepackage{cite}
\usepackage{amssymb}
\usepackage{amsfonts}
\usepackage{algorithm}
\usepackage{dsfont}
\usepackage{graphicx}
\usepackage{epsfig}
\usepackage{subfigure}
\usepackage{psfrag}
\usepackage{xcolor}
\usepackage{url}
\usepackage[colorlinks,linkcolor=black,urlcolor=black,anchorcolor=black,citecolor=black,hyperfootnotes=true]{hyperref}
\def\BibTeX{{\rm B\kern-.05em{\sc i\kern-.025em b}\kern-.08em
    T\kern-.1667em\lower.7ex\hbox{E}\kern-.125emX}}
\usepackage{balance}

\newtheorem{example}{Example}

\begin{document}
\title{Leveraging A Variety of Anchors in Cellular Network for Ubiquitous Sensing}
\author{Liang Liu, Shuowen Zhang, and Shuguang Cui
\thanks{Liang Liu and Shuowen Zhang (corresponding author) are with The Hong Kong Polytechnic University, Hong Kong SAR, China (e-mail: \{liang-eie.liu,shuowen.zhang\}@polyu.edu.hk).}
\thanks{Shuguang Cui is with The Chinese University of Hong Kong, Shenzhen, and the Future Network of Intelligence Institute (FNii), China (email: shuguangcui@cuhk.edu.cn).}}

\maketitle

\begin{abstract}
Integrated sensing and communication (ISAC) has recently attracted tremendous attention from both academia and industry, being envisioned as a key part of the standards for the sixth-generation (6G) cellular network. A key challenge of 6G-oriented ISAC lies in how to perform ubiquitous sensing based on the communication signals and devices. Previous works have made great progresses on studying the signal waveform design that leads to optimal communication-sensing performance tradeoff. In this article, we aim to focus on issues arising from the exploitation of the communication devices for sensing in 6G network. Particularly, we will discuss about how to leverage various nodes available in the cellular network as anchors to perform ubiquitous sensing. On one hand, the base stations (BSs) will be the most important anchors in the future 6G ISAC network, since they can generate/process radio signals with high range/angle resolutions, and their positions are precisely known. Correspondingly, we will first study the BS-based sensing technique. On the other hand, the BSs alone may not enable ubiquitous sensing, since they cannot cover all the places with strong line-of-sight (LOS) links. This motivates us to investigate the possibility of using other nodes that are with higher density in the network to act as the anchors. Along this line, we are interested in two types of new anchors - user equipments (UEs) and reconfigurable intelligent surfaces (RISs). This paper will shed light on the opportunities and challenges brought by UE-assisted sensing and RIS-assisted sensing. Our goal is to devise a novel 6G-oriented sensing architecture where BSs, UEs, and RISs can work together to provide ubiquitous sensing services.
\end{abstract}


\section{Introduction}
\IEEEPARstart{R}{ecently}, integrated sensing and communication (ISAC) has attracted tremendous attention \cite{Eldar19}. Particularly, in ``The ITU-R Framework for IMT-2030'' \cite{IMT2030}, ITU-R Study Group 5 identified ISAC as one of the six usage scenarios for the sixth-generation (6G) cellular network. Because communication is a standard and mature function in cellular network, one main challenge on the roadmap towards ISAC lies in how to effectively achieve the sensing functionality using the communication signals and devices available in the future 6G systems. This is the main topic that we aim to discuss about in this article.

There are several successful sensing systems in the world, e.g., radar sensing systems, Wi-Fi sensing systems, etc. A nature question thus arises - besides the commercial consideration, what are the technical advantages of the 6G-oriented sensing technique over the existing sensing techniques? The answer may lie in the choices of anchors. Anchors are imperative to localization. Specifically, we can leverage the radio signals to estimate the distances and angle-of-arrivals (AOAs) from the unknown positions of the targets to the known positions of the anchors. Then, based on the absolute state information of the anchors and the relative state information between the targets and the anchors, we can accurately localize the targets. In conventional sensing systems, radars and Wi-Fi access points act as the anchors, because they can actively emit radio signals and process received signals, and their positions are known. Based on the above philosophy, a straightforward strategy is to utilize the base stations (BSs) as anchors for sensing in 6G cellular network. Actually, most of the current works on 6G-oriented ISAC do rely on the BSs for performing communication and sensing functions. Inspired by the success of the BSs in providing ubiquitous communication services, people may expect the BSs to provide ubiquitous sensing services as well. However, this is not the truth because in practice, the density for deploying the BSs is not high enough to cover the targets everywhere with strong line-of-sight (LOS) channels.

\begin{figure*}[t]
\centering
\includegraphics[width=5.5in]{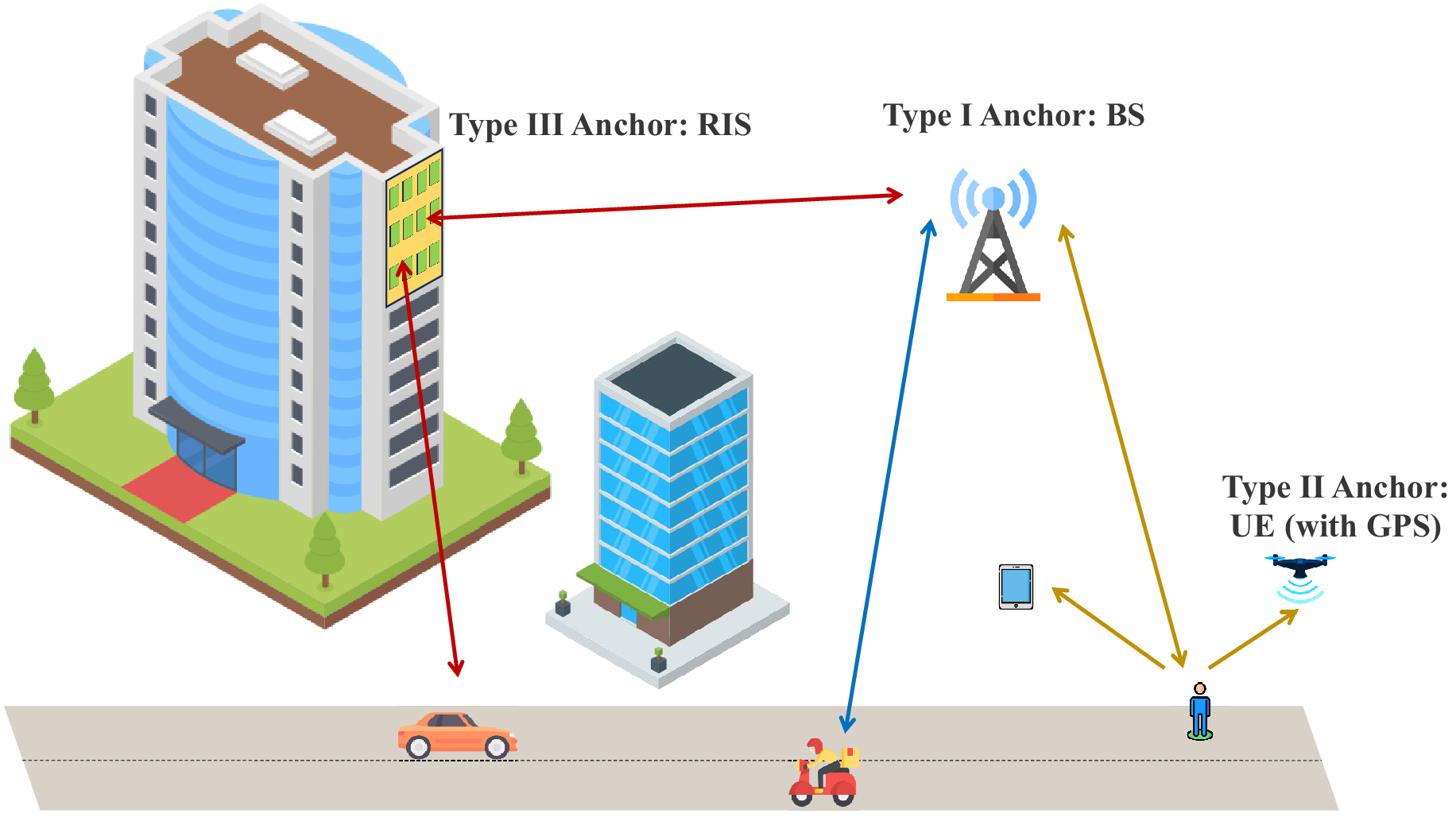}
\caption{An illustration of 6G-oriented sensing architecture with a variety of anchors such as the BS, the UE, and the RIS. Because the BS has a LOS path to the bike, it can act as an anchor to localize the bike alone. Because the BS and the two UEs all have LOS paths to the pedestrian, they can both serve as anchors to perform joint localization. Because merely the RIS has a LOS path to the vehicle, it acts as a passive anchor to reflect the signals from the vehicle to the BS, which can leverage these signals to localize the vehicle based on its relative position to the RIS.}
\label{fig1}
\end{figure*}

To overcome the above limitation, we propose to leverage a variety of nodes in the cellular network to expand the sensing regions of the BSs, as shown in Fig. \ref{fig1}. Along this line, two new types of anchors will be discussed in this article - user equipments (UEs) and reconfigurable intelligent surfaces (RISs). Our goal is to demonstrate the feasibility and the superiority of employing not only the BSs, but also the UEs and the RISs as anchors for high-performance sensing in 6G network. Via properly utilizing all possible nodes in the cellular network, we believe that the density of anchors will become the fundamental advantage of 6G-oriented sensing over the existing sensing techniques, which enables ubiquitous sensing.

At last, we want to emphasize that there are plenty of works focusing on how to utilize the 6G signals for achieving the optimal communication-sensing performance tradeoff. Our work is an early effort to discuss about how to leverage the 6G devices for better sensing performance.

\section{Types of Anchors in 6G Network for Sensing}\label{sec:anchors}
In this article, we will discuss about three types of anchors in the 6G cellular network - BSs, UEs, and RISs, as shown in Fig. \ref{fig1}. First, the BSs are the most powerful anchors and will form the foundation to 6G-oriented sensing. However, the key limitation for leveraging BSs as anchors lies in their low deployment density. This motivates us to investigate the possibility for employing the UEs and the RISs as anchors to probe the regions that cannot be seen clearly by the BSs alone.

On one hand, UEs, such as mobile phones and tablets, become more computationally powerful nowadays, and they have played an important role in myriad applications other than communications, including federated learning, mobile edge computing, etc. It is thus feasible and promising to leverage UEs with strong communication and computation capabilities as anchors for performing sensing tasks. The benefit to use UEs as anchors is quite straightforward - the density of the UEs is much higher than that of the BSs in the cellular network. In practice, after the BS emits the downlink radio signals over the air, the widely deployed UEs can capture different aspects of the radar cross section (RCS) of each target from different directions \cite{Haimovich08}. A main message from \cite{Haimovich08} is that thanks to the RCS diversity gain, it is beneficial to deploy the antennas at distributed locations such that there are always some antennas to receive strong echo signals from the targets. In 6G networks, this indicates that while it is sometimes hard to find BSs that can receive strong echo signals from a target, it is much easier to find some adjacent UEs that can receive strong echoes as anchors.

\begin{figure}[t]
\centering
\includegraphics[width=3.5in]{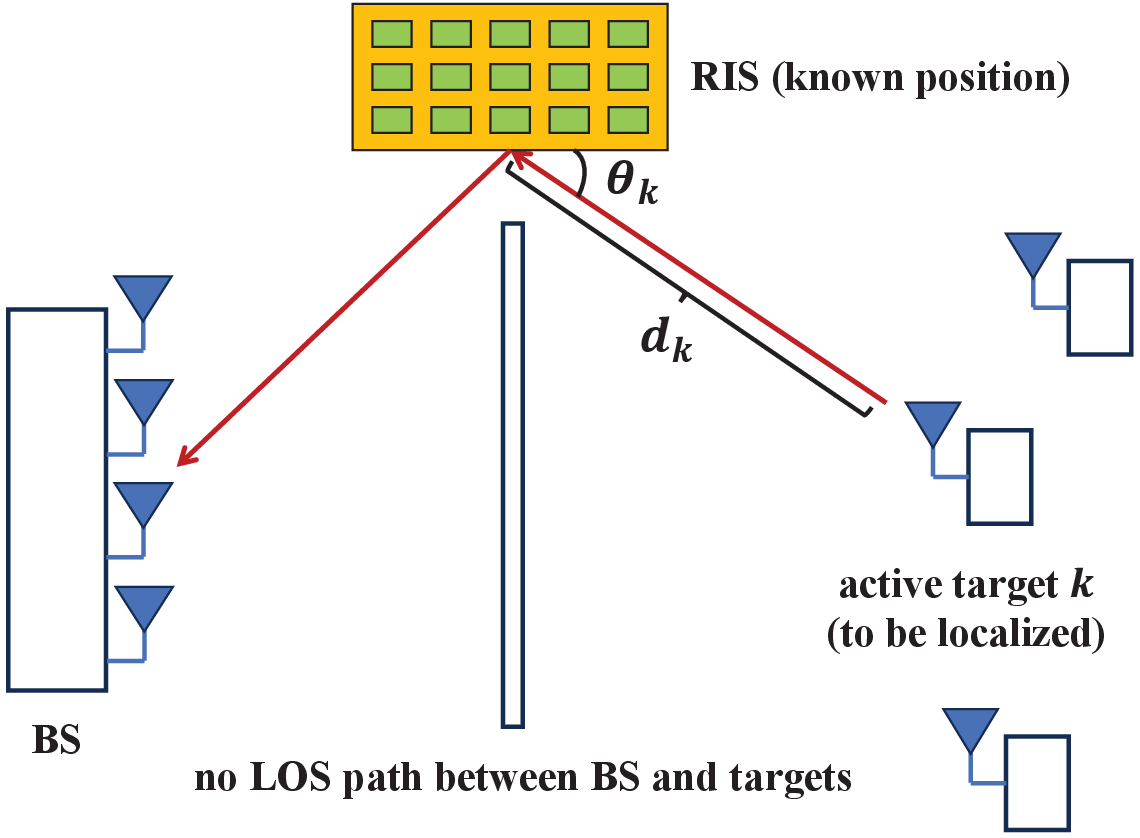}
\caption{An illustration of RIS-assisted location: An RIS assists the BS to localize the targets without LOS paths to it. Because the ranges and the AOAs from the targets to the BS are not contained in the signals over the target-RIS-BS paths, we have to localize the targets based on their ranges and AOAs to the passive anchor, i.e., the RIS.}
\label{fig2}
\end{figure}

On the other hand, RISs can act as passive anchors to extend the sensing region of the BSs. Specifically, when there lack LOS paths between the BS and the targets in a particular region, an RIS can be deployed at a known site with LOS paths to these targets and help reflect the signals from the targets to the BS, as shown in Fig. \ref{fig2}. In the above scenario, the RIS can act as a passive anchor due to the following reasons. First, the RIS can be an anchor because thanks to the LOS links between the targets and the RIS, it is theoretically possible to estimate each target's distance and AOA to the RIS and then localize the targets based on their relative states to the RIS with known position. Second, the RIS is a passive anchor because it cannot process its received signals from the targets to directly estimate the above useful information. Instead, it has to reflect the signals to the BS, which then indirectly estimates the relative states between the RIS and the targets. That is why an RIS with known position but without signal processing capability can act as a passive anchor to assist the BS to expand its sensing region.

To summarize, a promising 6G architecture for ubiquitous sensing is shown in Fig. \ref{fig1} - the BSs will still play the dominant role to probe the environment, while the widely deployed UEs may help the BSs to improve the sensing accuracy, and the RISs may assist the BSs to sense the regions that cannot be covered by them. In the rest of this article, we list some challenges for enabling BS-based sensing, UE-assisted sensing, and RIS-assisted sensing, and provide possible directions to tackle these challenges.

\section{BS-Based Sensing Strategy}
BSs form the foundation to wireless communication in cellular network. Therefore, it is not surprising that BSs will also be the basis for sensing in 6G cellular network. There are several desirable properties making BSs perfect anchors to perform sensing. First, the BSs are static, and their positions are precisely known. Second, the BSs can estimate the relative states of the targets with high resolution. On one hand, the BSs can emit the millimeter wave (mmWave) signals that are with large bandwidth. For example, according to 3GPP Release 15, the maximum bandwidth of 5G mmWave signals is 400 MHz, and the corresponding range resolution for sensing is $0.375$ meter. This is beneficial to range estimation. On the other hand, the antenna array at the BS is large thanks for the massive multiple-input multiple-output (MIMO) technique. This is beneficial to angle estimation.

One may argue that because the BSs play a similar role to radars in localization, the techniques for radar sensing and BS-based sensing would be almost the same. While, this is not the truth. In this section, we point out two unique issues about BS-based sensing - range/angle/Doppler estimation, and the possibility to perform networked sensing via large-scale BS cooperation.

\subsection{Range/Angle/Doppler Estimation via MIMO Orthogonal Frequency Division Multiplexing (OFDM) Signals}
Radar signals and cellular signals possess quite different waveforms, because they are designed for different purposes. Specifically, radar signals are designed to have ambiguity functions with steep and narrow main lobes such that a matched filter can accurately estimate the range and Doppler information from the echo signals. However, the OFDM signals in the cellular network are modulated, coded, and random signals for high-speed communication, and do not possess the above property of radar signals. As reported in \cite{Evers14}, if the radar-based matched filter is adopted for sensing over OFDM signals, the estimation performance can be very poor.

Similar to Wi-Fi sensing, we may enable 6G-oriented sensing based on channel state information (CSI). CSI captures how wireless signals travel through surrounding objects in time, frequency, and spatial domains. Because the existence of the targets can alter some properties of the propagation channels in a certain manner, we can extract range, angle, and Doppler information by analyzing CSI. In \cite{Liu20,Liu_ISAC_2022}, it was shown that in the MIMO-OFDM systems, the range/angle information can be estimated from the time-domain OFDM signals. To summarize, the main message here is that we can still extract the range/angle/Doppler information accurately from the MIMO-OFDM communication signals, but based on new signal processing algorithms, instead of the conventional one used for radar sensing.

\subsection{Networked Sensing}

\begin{figure}[t]
\centering
\includegraphics[width=3.5in]{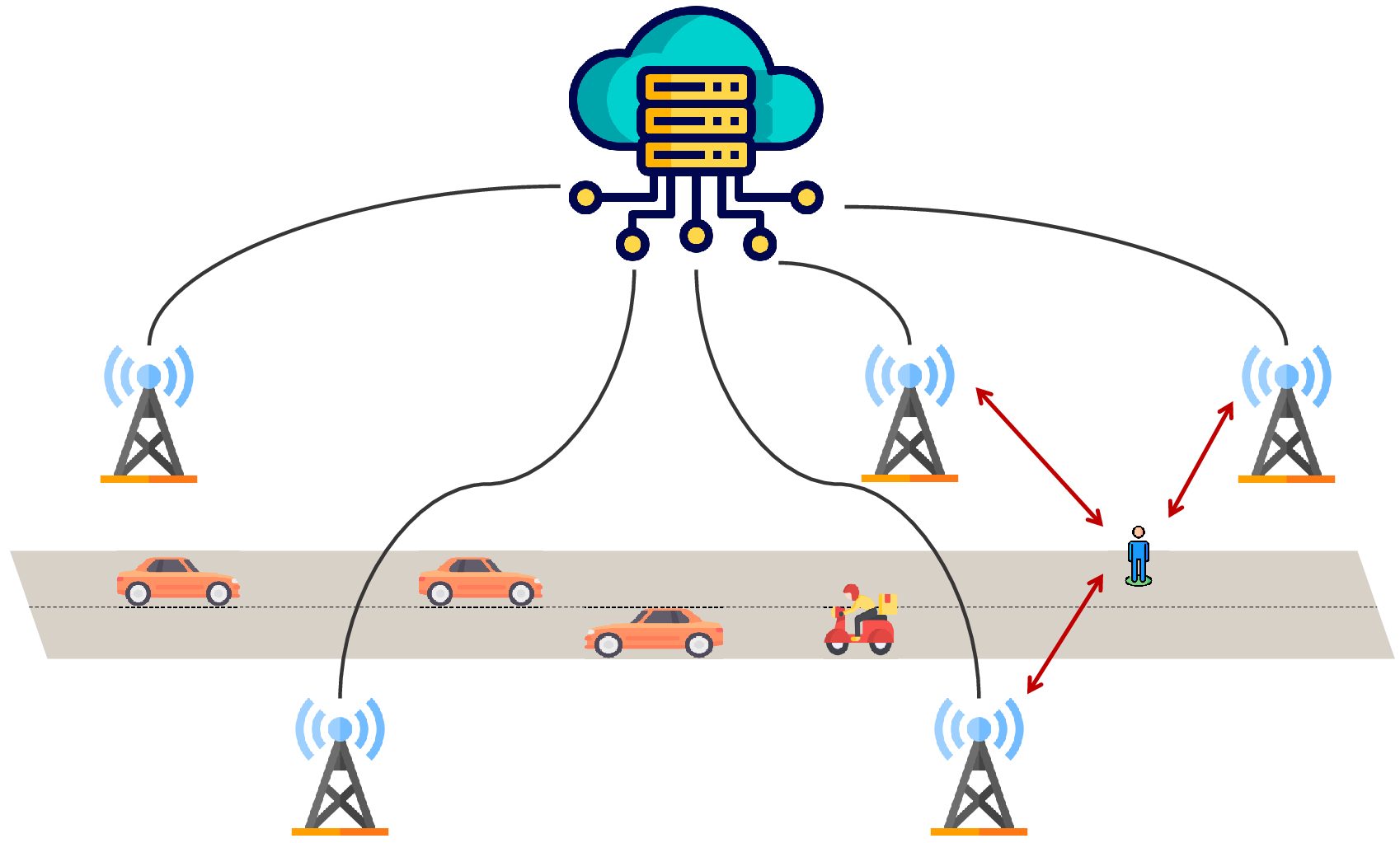}
\caption{An illustration of networked sensing: 5 BSs connected to the same cloud via the backhaul network are deployed along the road to monitor the traffic conditions. For example, the pedestrian can be jointly localized by three BSs with very high accuracy.}
\label{fig3}
\end{figure}

In wireless communication, because each user's signals can be received by multiple BSs, cooperative communication techniques, such as cloud radio access network, networked MIMO, etc., have been a hot topic for quite a while, where BSs collaboratively serve the users to mitigate the inter-cell interference. In wireless sensing, each target's signals can also be heard by multiple BSs. Therefore, the BSs can perform networked sensing \cite{Networked_sensing_2023} via sharing their local sensing information for better estimations of the environment, as shown in Fig. \ref{fig3}. Networked sensing is a unique advantage of 6G-oriented sensing over radar sensing, because radars usually work independently due to their sparse deployment and the absence of a globally unified standard.

Note that networked sensing system is a multi-source multi-target sensing system, and data association is a long-standing issue under such a system \cite{Data_Association_2007}. Specifically, to localize a particular target, we should utilize its echo signals received by multiple BSs for performing networked sensing. However, the echo signals from all the targets share the same signature, and it is difficult for the BSs to match each of its received echoes to the target that generates this echo signal. In the following, we provide an example to show that the data association issue may lead to detection of ghost targets that do not exist.

\begin{example}\label{example1}
Suppose there are three BSs with coordinates $(-35,0)$, $(50,0)$, and $(0,-45)$, and two targets with coordinates $(30,30)$ and $(-30,-30)$. If the range estimation is perfect, then BSs 1, 2, and 3 respectively have a distance set of $\{\sqrt{5125},\sqrt{925}\}$, a distance set of $\{\sqrt{1300},\sqrt{7300}\}$, and a distance set of $\{\sqrt{6525},\sqrt{1125}\}$, with the targets.
\end{example}

In the above example, if BSs 1, 2, and 3 respectively use the distances $\sqrt{5125}$, $\sqrt{1300}$, and $\sqrt{6525}$ for localizing target 1, and $\sqrt{925}$, $\sqrt{7300}$, and $\sqrt{1125}$ for localizing target 2, the coordinates of these two targets can be perfectly estimated as $(30,30)$ and $(-30,-30)$ by applying the trilateration method. However, if BSs 1, 2, and 3 respectively match $\sqrt{5125}$, $\sqrt{1300}$, and $\sqrt{1125}$ for localizing target 1, and $\sqrt{925}$, $\sqrt{7300}$, and $\sqrt{6525}$ for localizing target 2, the coordinates of these two targets will be estimated as $(30,-30)$ and $(-30,30)$. In this case, we define the false targets at $(30,-30)$ and $(-30,30)$ as the \emph{ghost targets} arising from the wrong data association solution.

The above example indicates that the BSs may detect ghost targets due to wrong data association. However, it is also worth noting that ghost targets do not always exist, as shown in the following example.

\begin{example}\label{example2}
Suppose the locations of the three BSs and the second target remain the same as in Example \ref{example1}, while the location of the first target is changed to $(30,20)$. If the range estimation is perfect, then BSs 1, 2, and 3 respectively have a distance set of $\{\sqrt{4625},\sqrt{925}\}$, a distance set of $\{\sqrt{800},\sqrt{7300}\}$, and a distance set of $\{\sqrt{5125},\sqrt{1125}\}$, with the targets.
\end{example}

In the above example, it can be shown that merely when BSs 1, 2, and 3 respectively use the distances $\sqrt{4625}$, $\sqrt{800}$, and $\sqrt{5125}$ for localizing target 1, and $\sqrt{925}$, $\sqrt{7300}$, and $\sqrt{1125}$ for localizing target 2, the coordinates of these two targets can be estimated as $(30,20)$ and $(-30,-30)$ by applying the trilateration method. When other data association solutions are used, there is no solution under the trilateration method. In other words, different from Example \ref{example1}, there is no ghost target issue if the BSs and the targets are distributed as in Example \ref{example2}. To summarize, the data association issue in networked sensing may or may not result in the ghost target issue, depending on the locations of the BSs and the targets.

The above data association issue has been investigated in \cite{Liu_ISAC_2022}, which sends two messages. First, we do not need to worry that the data association issue will fundamentally limit the sensing accuracy, because when the targets are randomly distributed in the network, almost surely just one data association solution can lead to a feasible target location solution under the trilateration method. Second, although merely one data association solution exists with probability one, how to find this correct solution is far from being simple. Some attempts for designing efficient data association algorithms have been made in \cite{Liu_ISAC_2022}. In the following, we provide one example to show the effectiveness of the data association algorithm proposed in \cite{Liu_ISAC_2022}.

\begin{figure}[t]
\centering
\includegraphics[width=3.5in]{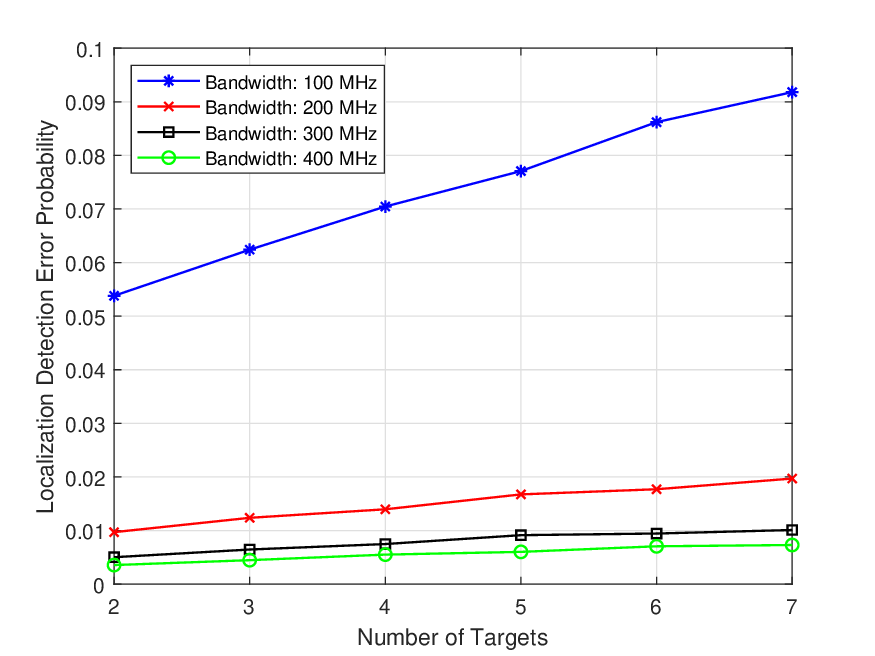}
\caption{Performance of the data association algorithm proposed in \cite{Liu_ISAC_2022}.}
\label{fig4}
\end{figure}

\begin{example}\label{example3}
Consider a networked sensing system consisting of $5$ BSs and $2$-$7$ targets. Fig. \ref{fig4} shows the localization detection error probability of the data association and localization algorithm proposed in \cite{Liu_ISAC_2022}, when the channel bandwidth ranges from $100$ MHz to $400$ MHz. Here, an detection error event for localizing the target is defined as the case that the estimated location does not lie within a radius of $1$ m from the true target location. It is observed that when the channel bandwidth is above $300$ MHz, the localization error probability is below $1\%$ when the number of targets ranges from $2$ to $7$.
\end{example}

Although the BSs are powerful anchors, they cannot cover anywhere with strong LOS links. Therefore, in the following, we introduce how to utilize the UEs and the RISs to expand the BSs' sensing regimes.

\section{UE-Assisted Sensing Strategy}
In this section, we demonstrate the possibility for using the UEs as the anchors to assist the BS in sensing. Specifically, under the UE-assisted sensing strategy, each BS can transmit the downlink signals to probe the environment, while the UEs receiving strong echo signals can opportunistically help localize the targets, as shown in Fig. \ref{fig1}. This is the so-called RCS diversity gain \cite{Haimovich08}. Note that instead of transmitting the echo signals, each UE merely needs to transmit a small amount of range, angle, and Doppler information over the feedback channel to the cloud to perform UE-assisted localization. Because of the high density of the UEs, it is quite likely that each target can find some nearby UEs with strong RCSs. However, there are several issues to implement UE-assist sensing, and dedicated techniques should be proposed to tackle these issues before we can reap the RCS diversity gain in practice. In the following, we list the challenges and the possible solutions for UE-assisted sensing.
\subsection{Timing Offsets (TOs) Among Asynchronous Anchors}
In practice, it is impossible to perfectly synchronize the BS and the UEs under the UE-assisted sensing scheme. In this case, the propagation delay from the BS to a target to an UE estimated from the echo signal is actually the superposition of the true propagation delay and the TO between the asynchronous BS and UE. Therefore, we need to propose efficient methods to mitigate the effect of TO on range-based localization under UE-assisted sensing.

It is worth noting that mitigating the effect of TO on communication is not hard, as long as the length of the cyclic prefix (CP) in OFDM symbols is sufficiently large. In this case, inter-symbol interference (ISI) still lies in the CP of all users' received signals in asynchronous systems, and after removing the CP, TO just adds linear phases in frequency-domain channels of each user, which can be simply compensated for via channel estimation without knowing TOs \cite{Morelli07}. However, TO estimation is necessary for localization, because range estimation is affected by TOs.

There are several ways to mitigate the effect of TO on range estimation. First, if the LOS path between the BS and an UE is available, we can first calculate the true delay from the BS directly to the UE based on their locations, and then utilize the UE's received signal over the LOS path to estimate the effective propagation delay. The time difference between the true delay and the effective delay will be the TO between the BS and the UE \cite{Xianzhen}. Second, if the LOS path between the BS and an UE is not available, we should jointly estimate the TO and target locations, i.e., via solving the maximum-likelihood (ML) problem.

\subsection{Erroneous Position Information of Anchors}
Anchors are conventionally defined as the nodes whose state information is perfectly known. However, different from the static radars and BSs, the UEs can move and their positions have to be dynamically estimated by Global Positioning System (GPS). Such estimations are subject to unknown errors, which can be quite large sometimes. If some UEs with quite erroneous position information are selected as the anchors, the sensing accuracy can be very low. Therefore, when multiple UEs can receive strong echo signals from the targets, we have to propose efficient methods to find the UEs with very accurate position information for acting as the anchors.

In the literature, some works have been done for localization with erroneous anchor position information \cite{Anchor_14}. In these works, the distribution of the anchor position errors is assumed to be known, based on which the ML problem for localization is formulated. It turns out that such an ML problem is a weighted sum-residue minimization problem, where higher weights are assigned to the residues of the anchors with smaller error power. In other words, the anchors with favorable distribution of position errors will play more important roles in estimating the locations of the targets. However, in practice, the distribution of the anchor position errors is usually not available. We should design UE selection algorithms that do not rely on any prior information about UE position errors.

We may achieve the above goal via utilizing the outlier detection technique \cite{Hodge04}. Specifically, the sensing information obtained by the UEs with quite erroneous position information can be treated as the outlier. For example, in \cite{Xianzhen}, an outlier-based iterative UE selection algorithm is proposed, where one UE with quite erroneous position information is found at each iteration. The basic idea is as follows. At the beginning, we use all the UEs as the anchors to localize the targets. The corresponding estimation residue should be recorded. In the first iteration of the UE selection algorithm, if the removal of one UE can result in the maximal localization residue reduction compared to the removal of any other UE, then this UE will be treated as an UE with quite erroneous position information and removed from the anchor set. We can perform this iteration by iteration, until the removal of any UE will not lead to notable localization residue reduction. Then, we can claim that all the UEs in the anchor set are UEs with accurate position information.

\begin{figure}[t]
\centering
\includegraphics[width=3.5in]{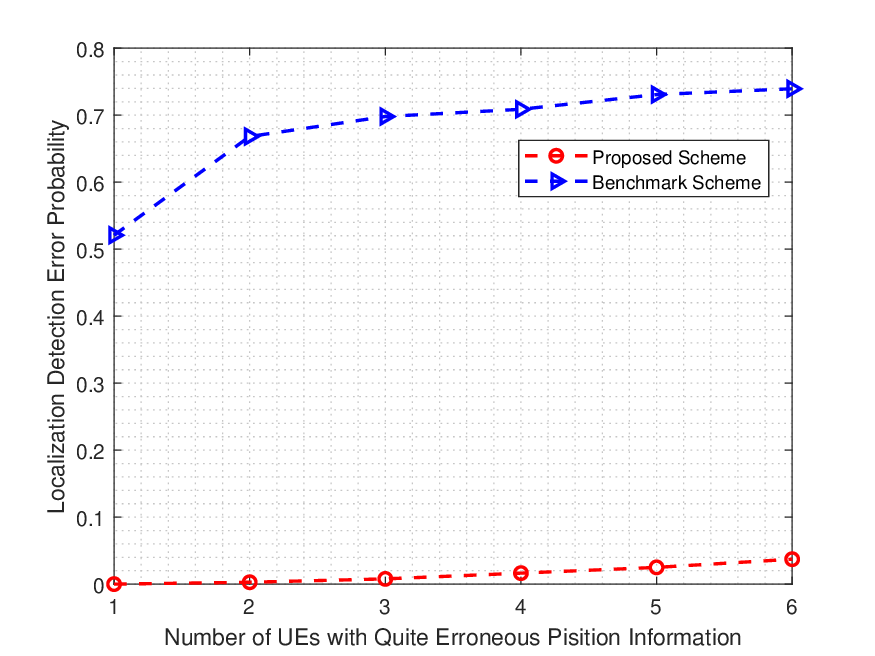}
\caption{An illustration of localization accuracy when position information of the anchors is not always precise.}
\label{fig5}
\end{figure}

\begin{example}\label{example4}
Consider an UE-assisted localization example, where one BS works with several UEs to localize one target. Among the UEs, we assume that $5$ are with accurate position information, while $1-6$ UEs are with quite erroneous position information. Fig. \ref{fig5} shows the localization detection error probability when the above UE selection algorithm from \cite{Xianzhen} is applied to find the UEs with quite erroneous position information and remove them from the anchor set. Moreover, the strategy to use all the UEs as anchors is the benchmark scheme. It is observed that selecting the UEs with accurate position information as anchors is indeed a necessary step for UE-assisted localization.
\end{example}

\section{RIS-Assisted Sensing Strategy}
Recently, the RIS has become a promising technology for high-speed communication in the 6G network. The great potential for using RISs to improve the network throughput has been demonstrated by a large body of research works. In this section, we aim to show that RISs are beneficial not only to communication, but also to sensing. Note that BSs and UEs are all active anchors, in the sense that they can transmit/receive radio signals for the sensing purpose. However, as explained in Section \ref{sec:anchors}, the RISs are passive anchors, because they can merely reflect their received signals, instead of processing them. Such a passive nature brings challenges to RIS-assisted sensing.

\subsection{Range and AOA Estimation with Passive Anchors}
Conventional anchors, such as radars and BSs, are all active anchors, which can exploit their received signals to localize the targets if there exist LOS paths among them. However, RISs are passive devices that can merely reflect the signals. Thereby, as shown in Fig. \ref{fig2}, the main challenge for RIS-assisted sensing lies in how to estimate each target's range and AOA to the RIS, if the RIS cannot process its received signals. The only possible solution is to let the BS estimate these useful information based on its received signals over the target-RIS-BS links. This is theoretically feasible, because the signals received by the BS over the target-RIS-BS links are functions of the signals received by the RIS over the target-RIS links, which are functions of the targets' distances and AOAs to the RIS.

Specifically, we can estimate the propagation delays from the targets to the RIS as follows \cite{Wang22}. First, we can measure the propagation delay from each target to the RIS to the BS based on the signal received by the BS over the target-RIS-BS path. Second, because the positions of the BS and the RIS are known, we can calculate the propagation delay between the RIS and the BS, based on the distance between them. Last, the difference between the above two values will be the propagation delay from a target to the RIS.

The next topic is how to estimate the AOAs from the targets to the RIS based on the signals received by the BSs. Note that if we apply the multiple signal classification (MUSIC) algorithm directly on the spatial-domain signal vector received across all the antennas of the BS, we can merely estimate the AOA from the RIS to the BS, because the incident signals to the BS is from the RIS. However, our interest is on the AOAs from the targets to the RIS, which are useful to localize the targets. Interestingly, our recent work \cite{Wang23} showed an amazing result - if the MUSIC algorithm is applied on a properly designed temporal-domain received signal vector of the BS, we are still able to estimate the AOAs from the targets to the RIS, instead of the AOA from the RIS to the BS. Here, the novelty is to dynamically change the RIS reflection coefficients over time so as to create the temporal-domain received signal vector of the BS. More information about how to control the RIS and how to construct the temporal-domain received signals can be found in \cite{Wang23}.

\begin{figure}[t]
\centering
\includegraphics[width=3.5in]{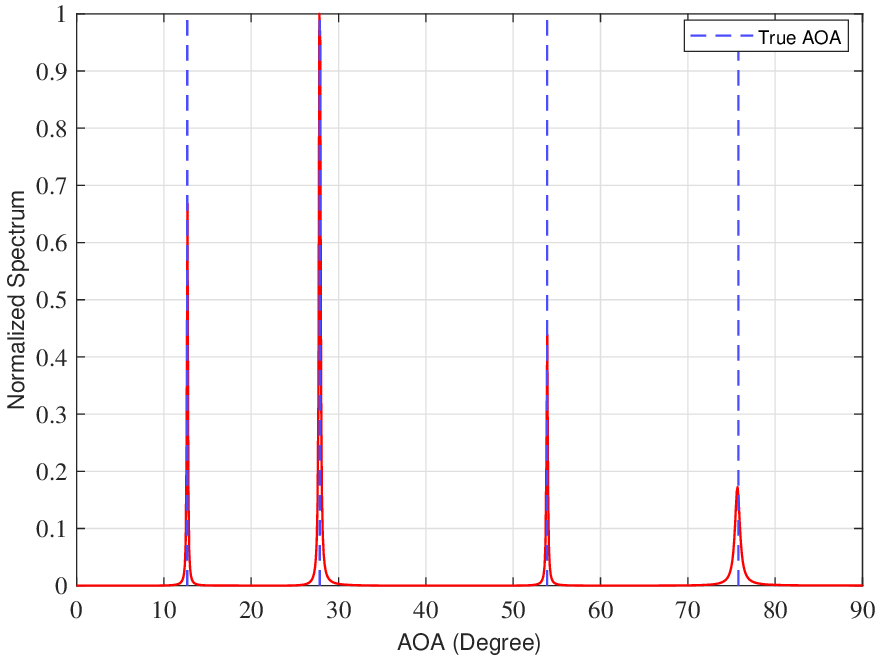}
\caption{An illustration of the accuracy of AOA estimation under RIS-assisted localization.}
\label{fig6}
\end{figure}

\begin{example}\label{example4}
Consider an RIS-assisted localization example, where $4$ targets are at sites with LOS paths to a 64-element RIS. The AOAs from the $4$ targets to the RIS are $12.6728^\circ$, $27.8523^\circ$, $53.8847^\circ$, and $75.7906^\circ$. Fig. \ref{fig6} shows the normalized spectrum when the MUSIC algorithm is applied to the carefully designed temporal-domain signals of the BS. It is observed that all the $4$ AOAs are accurately estimated.
\end{example}

\subsection{Sequentially Updating RIS Reflection Coefficients for Better Localization Performance}
The key to make the MUSIC algorithm work for estimating the AOAs from the targets to the RIS lies in the construction of the temporal-domain signals at the BS side, which critically relies on the dynamic design of the RIS reflection coefficients over time. This is because if a better RIS beamforming solution can be obtained at a time slot, the signals emitted by the active targets in the next time slot can be reflected to the BS with more focused power for better localization performance. Along this line, the active beamforming technique proposed in \cite{Yu22} can be applied to sequentially optimize RIS reflection strategies based on BS's historical observations. One notable challenge of this approach is that the dimension of historical data changes over time. It is known that recurrent neural network (RNN) is powerful to process sequential date. The active beamforming technique proposed in \cite{Yu22} is based on RNN and promising to solve the RIS reflection coefficient design problem.

\section{Conclusions}
In this article, we have discussed about the possibility to leverage a variety of anchors, including BSs, UEs, and RISs, for ubiquitous sensing in 6G cellular network. Specifically, BSs will be the most important anchors, and the UEs and the RISs can assist the BSs to expand their sensing regions. We have listed and provided possible solutions to the challenges for BS-based sensing (including range, angle, and Doppler estimation methods as well as the data association issue for networked sensing), UE-assisted sensing (including effect of timing offset on range estimation as well as effect of erroneous anchor position information), and RIS-assisted sensing (including range, angle, and Doppler estimation via passive anchors as well as adaptive RIS reflection strategies). We believe that via properly utilizing various nodes in cellular network, it is promising to transform the world's largest wireless network into the world's largest sensing network in the future.

\section{Acknowledgment} 
The work was supported in part by the National Key R$\&$D Project of China under Grant No. 2022YFB2902800; by the Research Grants Council, Hong Kong, China, with Grant No. 15203222 and Grant No. 15230022; by the National Natural Science Foundation of China with Grant No. 62101474 and Grant No. 62293482; by the Basic Research Project No. HZQB-KCZYZ-2021067 of Hetao Shenzhen-HK S$\&$T Cooperation Zone; by the National Key R$\&$D Program of China with grant No. 2018YFB1800800; by the Shenzhen Outstanding Talents Training Fund 202002; by the Guangdong Research Projects No. 2017ZT07X152 and No. 2019CX01X104; by the Guangdong Provincial Key Laboratory of Future Networks of Intelligence (Grant No. 2022B1212010001); by the Shenzhen Key Laboratory of Big Data and Artificial Intelligence with Grant No. ZDSYS201707251409055; and by the Key Area R$\&$D Program of Guangdong Province with Grant No. 2018B030338001.

\begin{IEEEbiographynophoto}{Liang Liu} (Senior Member, IEEE) received the Ph.D. degree from the National University of Singapore, Singapore, in 2014. He is currently an Assistant Professor with the Department of Electrical and Electronic Engineering, The Hong Kong Polytechnic University. He was a recipient of the 2021 IEEE Signal Processing Society Best Paper Award, the 2017 IEEE Signal Processing Society Young Author Best Paper Award, the Best Student Award of of 2022 IEEE International Conference on Acoustics, Speech, and Signal Processing (ICASSP), and the Best Paper Award of the 2011 International Conference on Wireless Communications and Signal Processing. He was recognized by Clarivate Analytics as a Highly Cited Researcher in 2018. He is an Editor of IEEE TRANSACTIONS ON WIRELESS COMMUNICATIONS. He is a co-author of the book ``Next Generation Multiple Access'' published at Wiley-IEEE Press.
\end{IEEEbiographynophoto}

\begin{IEEEbiographynophoto}{Shuowen Zhang} (Member, IEEE) received the Ph.D. degree from the NUS Graduate School for Integrative Sciences and Engineering (NGS), National University of Singapore, in January 2018, under the NGS Scholarship. Since 2020, she has been with the Department of Electrical and Electronic Engineering, The Hong Kong Polytechnic University, where she is currently an Assistant Professor. Her current research interests include intelligent reflecting surface-aided communications, unmanned aerial vehicle (UAV) communications, and multiuser multiple-input multiple-output (MIMO) communications. She served as a Guest Editor for IEEE Journal on Selected Areas in Communications. She currently serves as the IEEE Communications Society Asia/Pacific Region Board WICE Vice Chair, and the ACM/IEEE N2Women Mentoring Co-Chair. She is the sole recipient of the Marconi Society Paul Baran Young Scholar Award, 2021, and recipient of the IEEE Communications Society Young Author Best Paper Award, 2022.
\end{IEEEbiographynophoto}

\begin{IEEEbiographynophoto}{Shuguang Cui} (Fellow, IEEE) received his Ph.D in Electrical Engineering from Stanford University, California, USA, in 2005. He is now a Distinguished Presidential Chair Professor at The Chinese University of Hong Kong, Shenzhen. His current research interest is data driven large-scale information analysis and system design. He was selected as the Thomson Reuters Highly Cited Researcher and listed in the Worlds' Most Influential Scientific Minds by ScienceWatch in 2014. He was the recipient of the IEEE Signal Processing Society 2012 Best Paper Award. He has been serving as the area editor for IEEE Signal Processing Magazine, and associate editors for IEEE Transactions on Big Data, IEEE Transactions on Signal Processing, and IEEE Transactions on Wireless Communications. He is an IEEE Fellow and ComSoc Distinguished Lecturer. In 2023, he won the IEEE Marconi Best Paper Award, got elected as a Fellow of both Royal Soceity of Canada and Canadian Academy of Engineering, and starts to serve as the Editor-in-Chief for IEEE Transactions on Mobile Computing.
\end{IEEEbiographynophoto}

\end{document}